\documentstyle[12pt]{article}

\newcommand{\be}{\begin{equation}}
\newcommand{\ee}{\end{equation}}
\newcommand{\dd}{\partial}
\newcommand{\bea}{\begin{eqnarray}}
\newcommand{\eea}{\end{eqnarray}}

\begin{document}
\baselineskip .25in
\newcommand{\numero}{hep-th/9512126, SHEP 95/40}

\newcommand{\titre}{On Non-Abelian Duality in Sigma Models}
\newcommand{\auteura}{Noureddine Mohammedi}
\newcommand{\place}{Department of Physics\\University of
Southampton\\
Southampton SO17 1BJ \\ U.K. }
\newcommand{\beq}{\begin{equation}}
\newcommand{\eeq}{\end{equation}}

\newcommand{\abstrait}
{A method for implementing non-Abelian duality on string backgrounds
is given. It is shown that a direct generalisation of the familiar Abelian
duality induces an extra local symmetry in the gauge invariant theory. The
non-Abelian isometry group is shown to be enlarged to a non-semi-simple
group. However, upon eliminating the gauge fields to obtain the dual theory
the new algebra does not close. Therefore the gauge fixing procedure
becomes problematic. The new method proposed here avoids these issues and
leads to a dual theory in the proper sense of duality.}
\begin{titlepage}
\hfill \numero  \\
\vspace{.5in}
\begin{center}
{\large{\bf \titre }}
\bigskip \\ by \bigskip \\ \auteura
\,\,\footnote{e-mail: nouri@hep.phys.soton.ac.uk}
    \bigskip \\ \place \bigskip \\

\vspace{.9 in}
{\bf Abstract}
\end{center}
\abstrait
 \bigskip \\
\end{titlepage}
\newpage
\section{Introduction}
Target space duality has been the focus of many investigations in the
last decade
(see \cite{amit} for a review and further references).
This is due to its relevance to the understanding
of the structure of the string backgrounds (or moduli space)
on which compactification
takes place. This duality connects two conformal backgrounds
which are, a priori, totally different. The simplest examples of this
duality are the $O(d,d,R)$ Narain \cite{narain}
transformation relating all possible
toroidal compactifications in $d$ dimensions. Generalisations of
Narain's transformations have also been found for backgrounds possessing
$d$ Abelian isometries \cite{thelot}.
This type of duality is refered to as the Abelian
duality. The duality is non-Abelian if the backgrounds isometries are
non-Abelian.
\par
The Abelian duality is better understood at the level of the Lagrangian
defining the low energy of string theory\footnote
{Duality has also been treated in terms of
canonical transformations \cite{lozano}.}
\cite{verlinde}. The dual non-linear
sigma model is obtained by gauging the $U(1)$ isometries of this Lagrangian
and at the
same time constraining, by means of Lagrange multipliers, the field
strength of each $U(1)$ gauge field to vanish. Integrating out these
Lagrange multipliers and fixing the gauge invariance yields the original
model. On the other hand, integrating over the gauge fields and keeping
the Lagrange multipliers results in the dual sigma model.
If the original backgrounds are conformal backgrounds (i.e. they satisfy
the vanishing of the beta functions conditions) then the dual backgrounds
are also conformal and form consistent backgrounds on which the string
propagates \cite{busher}.
\par
The generalisation of the above construction to backgrounds possessing
non-Abelian isometries could, in principle, be carried out in
exact analogy with the Abelian case \cite{delaossa}.
However, this straightforward
generalisations proves to be problematic \cite{israel}.
In particular, the dual
backgrounds do not, in general, satisfy the vanishing of the beta
functions conditions.
\par
In this letter we show that the Lagrange multiplier term needed
for non-Abelian duality cause the dual Lagrangian to have an
extra symmetry which, when combined
with the non-Abelian isometry group, forms a non-semi-simple
closed algebra. This new algebra does not close, however, when the
gauge fields are integrated out. This fact makes gauge fixing
difficult.
\par
This new symmetry arise because the Lagrange multipliers carry now
internal indices. A similar symmetry has appeared in the context
of gauge theories in four dimensions \cite{me1}.
The method we propose
for non-Abelian duality uses Lagrange multiplier carrying no
gauge group indices.
The implementation of this method is a natural one which stems from
the mathematical formalism of gauging isometries in a general
non-linear sigma model \cite{ian,chris}.
\par
We start, in section one, by stating the general formalism of
gauging isometries in non-linear sigma model. The usual way of
finding the dual theory is also explained. This is in direct analogy
with the Abelian case. The new symmetry is revealed in section three
and the new method is given.

\section{The Abelian analogy}

We start this section by giving a summary of gauging Abelian
and non-Abelian isometries
in a general sigma model.
The action for the general (ungauged) bosonic two-dimensional non-linear
sigma model is given by
\be
S(\varphi)=\int {\mathrm {d}}^2x \sqrt{\gamma}\left(
\gamma^{\mu\nu}g_{ij}\left(\varphi\right)\dd_\mu\varphi^i\dd_\nu\varphi^j
+\widehat{\epsilon}^{\mu\nu}b_{ij}\left(\varphi\right)
\dd_\mu\varphi^i\dd_\nu\varphi^j + D\left(\varphi\right)R^{(2)}\right)\,\,\,.
\ee
In this equation $\gamma_{\mu\nu}$ is the metric on the two-dimensional
world sheet, $\gamma$ is its determinant, $R^{(2)}$ is the scalar
curvature and $\widehat\epsilon^{\mu\nu}=\epsilon^{\mu\nu}/\sqrt\gamma$ is
the alternating tensor. The metric $g_{ij}$, the anti-symmetric tensor
$b_{ij}$ and the dilaton field $D$ correspond to the set of massless
modes of the associated string theory.
\par
The sigma model Lagrangian is manifestly invariant under global
reparametrisation of the target space. However, the only global symmetries
suitable for gauging are those for which the metric remains form invariant.
Such symmetries form a Lie group $G$, the isometry group of the metric.
A general infinitesimal global isometry is given by the transformation
\be
\delta \varphi^i=\alpha^a K^i_a\left(\varphi\right) \,\,\,.
\label{phi}
\ee
The generator of this transformation are
$K_a=K^i_a\dd_i$ and they satisfy the
Lie algebra of $G$
\be
\left[K_a\,\,,\,\,K_b\right]=f^c_{ab}K_c \,\,\,\,\,\,
{\mathrm{or}}\,\,\,\,\,\,
K^i_a\dd_iK_b^j-K^i_b\dd_iK^j_a=f^c_{ab}K^j_c\,\,\,,
\ee
where $f^a_{bc}$ are field-independent structure constants.
\par
The procedure for obtaining the dual sigma model consists in gauging the
above global symmetry and constraining the gauge field strength to vanish
through the addition of a Lagrange multiplier.
It is straightforward to gauge the first term in the Lagrangian. This is
achieved through the familiar mininal coupling
\be
\dd_\mu\varphi^i\,\,\,\,\,\, \longrightarrow\,\,\,\,\,
D_\mu=\dd_\mu\varphi^i + A^aK^i_a
\ee
and requiring that under  (\ref{phi}),
with $\alpha^a$ now a local function,
\be
\delta A^a_\mu=-\dd_\mu\alpha^a - f^a_{bc}\alpha^b A^c_\mu\,\,\,\,,
\ee
where $K^i_a$ satisfy the Killing vector condition
\be
K_a^k\dd_k g_{ij} + g_{kj}\dd_i K^k_a + g_{ik}\dd_j K^k_a =0 \,\,\,.
\ee
Gauging the second term of the Lagrangian requires, however,
a special treatment. Under the global isometry transformation,
this term is invariant provided that
\be
K^k_a\dd_k b_{ij}+ b_{kj}\dd_i K^k_a + b_{ik}\dd_j K^k_a
=\dd_i L_{aj}-\dd_j L_{ai}
\label{L}
\ee
for some vector $L_{ai}$. Because of this last equation,
the minimal coupling used to gauge the first term does not
lead to a gauge invariant action. The gauged invariant theory is found through
the use of Noether method and we have \cite{ian,chris}
\bea
S\left(\varphi, A\right)&=&\int{\mathrm{d}}^2x \sqrt{\gamma}
\left[\gamma^{\mu\nu}g_{ij}D_\mu\varphi^i D_\nu\varphi^j
+\widehat\epsilon^{\mu\nu}\left(b_{ij}\dd_\mu\varphi^i \dd_\nu\varphi^j
-2A^a_\mu c_{ai}\dd_\nu\varphi^i -d_{ab}A^a_\mu A^b_\nu\right)
\right.\nonumber\\
&+&\left. D(\varphi)R^{(2)}\right]
\eea
with $d_{ab}$ anti-symmetric and where
\bea
c_{ai}&=&b_{ij}K^j_a + L_{ai}\nonumber\\
d_{ab}&=&c_{ia}K^i_b\,\,\,.
\eea
The action is then gauge invariant provided that
\bea
K_b^j\dd_j c_{ai} + c_{aj}\dd_iK^j_b&=&-f^e_{ab}c_{ei}\nonumber\\
K^i_c\dd_id_{ab}&=& f^e_{ca}d_{eb}-f^e_{cb}d_{ea}\,\,\,\,
\label{gaugecond}
\eea
with $L_{ai}$ as defined in (\ref{L}). There are other ways of writing
the gauge invariance conditions showing explicitly the
anti-symmetry of $d_{ab}$ and involving only $L_{ai}$ \cite{ian}.
\newline
Finally, the dilaton term is gauge invariant provided that
$K^i_a\dd_iD=0$.
\par
If one follows the analogy with the Abelian duality then
one considers the following action \cite{delaossa}
\be
S\left(\varphi,A,\lambda\right)=S\left(\varphi,A\right) +
\int{\mathrm{d}}^2x
\sqrt{\gamma}\widehat\epsilon^{\mu\nu}F^a_{\mu\nu}\lambda_a\,\,\,,
\label{gaugaction}
\ee
where
$F^a_{\mu\nu}=\dd_\mu A_\nu^a -\dd_\nu A^a_\mu- f^a_{bc}A_\mu^b A_\nu ^c$
and for gauge invariance the Lagrange
multiplier $\lambda_a$ transforms as
\be
\delta \lambda_a = - f^c_{ab} \alpha^b\lambda_c\,\,\,.
\ee
Integrating over the Lagrange multiplier $\lambda_a$ leads to
$F_{\mu\nu}^a=0$ which is solved by $A^a_\mu=e^a_i(X)\dd_\mu X^i$,
where $e^a_i(X)$ are vielbeins satisfying the Maurer-Cartan conditions.
Choosing a gauge in which $A^a_\mu=0$
leads then to the
original action $S(\varphi)$.
Now keeping the Lagrange multiplier and integrating out
the gauge fields, instead, leads to the dual theory.
The equations of motion
for the gauge field (which is the same as doing the Gaussian integral)
is
\be
A^c_\rho = \widetilde M^{ca}_{\rho\mu}
\left[\left(\widehat\epsilon^{\mu\nu}c_{ai} -
\gamma^{\mu\nu}g_{ij}K^j_a\right)\dd_\nu\varphi^i
-\widehat\epsilon^{\mu\nu}\dd_\nu\lambda_a\right]
\,\,\,,
\label{gaugefield}
\ee
where $\widetilde M^{\mu\nu}_{ab}$ is defined by
\bea
M^{\mu\nu}_{ab} &=& \gamma^{\mu\nu}g_{ij}K^i_aK^j_b
-\widehat\epsilon^{\mu\nu}\left(d_{ab}+f^e_{ab}\lambda_e\right)
\nonumber\\
M^{\mu\nu}_{ab}\widetilde M^{bc}_{\nu\rho} &=& \delta^\mu_\rho
\delta^a_c\,\,\,.
\eea
Substituting for $A^a_\mu$ in
(\ref{gaugaction}) we get the following action
\bea
S\left(\varphi,\lambda\right) &=& S\left(\varphi\right) +
\int{\mathrm{d}}^2x\sqrt{\gamma}\left\{
-\widetilde M^{ad}_{\beta\alpha}\left(\widehat\epsilon^{\alpha\mu}
\widehat\epsilon^{\beta\nu} c_{di}c_{aj} +
\gamma^{\alpha\mu}\gamma^{\beta\nu}g_{il}g_{jk}K^l_dK^k_a
\right.\right.\nonumber\\
&-& \left.2 \widehat\epsilon^{\alpha\mu}\gamma^{\beta\nu} c_{di}g_{jk}K^k_a
\right)\dd_\mu\varphi^i\dd_\nu\varphi^j
-\widetilde M^{ac}_{\sigma\rho}\widehat\gamma^{\sigma\nu}\widehat
\epsilon^{\rho\mu}\dd_\nu\lambda_a\dd_\mu\lambda_c
\nonumber\\
&+& \left.2 \widetilde M^{ad}_{\rho\alpha}\left(\widehat\epsilon^{\rho\nu}
\widehat\epsilon^{\alpha\mu} c_{di}
-\widehat\epsilon^{\alpha\nu}
\gamma^{\rho\nu} g_{ij}K^j_d\right)\dd_\nu\lambda_a\dd_\mu\varphi^i
\right\}\,\,\,.
\label{dualaction}
\eea
Naively one would say that there are $\dim (G)$ gauge parameters $\alpha^a$
which would be used to eliminate $\dim (G)$ fields from this action and one
would then obtain a dual sigma model having the same number of fields as
the original theory. However, the action $S(\varphi, A, \lambda)$ as given
in (\ref{gaugaction}) has in fact an extra symmetry as shown below.

\section{The new method}

It is easy to see that by anti-symmetry the action (\ref{gaugaction})
is, in addition to its gauge invariance,  also invariant under
the local transformations
\bea
\widetilde\delta \lambda_a &=& \widehat\epsilon
^{\mu\nu}f^e_{ab}\xi_e F_{\mu\nu}^b\nonumber\\
\widetilde\delta \varphi^i &=& \widetilde\delta A_\mu^a = 0 \,\,\,,
\eea
where $\xi_a$ is the local gauge parameter.
Therefore the gauge invariance group has been enlarged due to the
addition of the Lagrange multiplier. The two local gauge transformation
$\delta_\alpha$ and $\widetilde\delta_\xi$ form the following closed
algebra
\bea
\left[\delta_\alpha\,,\,\delta_\beta\right]
&=&\delta_\gamma
\,\,\,\,\,,\,\,\,\,\,
\left[\delta_\alpha\,,\,\widetilde\delta_\omega\right]=
\widetilde\delta_\rho\nonumber\\
\left[\widetilde\delta_\alpha\,,\,
\widetilde\delta_\beta\right]&=&0\,\,\,\,,
\eea
with $\gamma^a=f^a_{bc}\alpha^b\beta^c$ and
$\rho_a=f^c_{ab}\alpha^b\omega_c$.
\par
After eliminating the gauge fields by their
equations of motion this is still
a symmetry of $S\left(\varphi,\lambda\right)$ where $F^a_{\mu\nu}$
is now the field strength corresponding to $A^a_\mu$ as given
in (\ref{gaugefield}).
We have explicitly checked that this is indeed the case. However
the elimination of the gauge fields causes the above algebra not
to close. This is a familiar issue in supergravity theories.
Therefore one does not know how to fix the gauge invariance of the action
$S\left(\varphi,\lambda\right)$.
\par
One way out of this problem is to consider, instead of (\ref{gaugaction}),
the following Lagrange multiplier term
\be
I\left(\varphi,A,\lambda\right)=S\left(\varphi,A\right) +
\int{\mathrm{d}}^2x
\sqrt{\gamma}\widehat\epsilon^{\mu\nu}F^a_{\mu\nu}
\Gamma_a
\lambda\,\,\,.
\label{newlag}
\ee
The Lagrange multiplier $\lambda$ does not carry now a gauge index
and the extra gauge symmetry found previously is not present
($\delta\lambda=\widetilde\delta\lambda=0$).
The new quantity $\Gamma_a(\varphi)$ is a function which satisfies
\be
K^i_a\dd_i\Gamma_b=f^c_{ab}\Gamma_c\,\,\,.
\ee
This equation is necessary for gauge invariance. This Lagrange multiplier
term reduces to the usual one when considering the Abelian case.
\par
Another property of the the new Lagrange multiplier term can be seen
by performing an integration by parts in (\ref{newlag})
\be
I\left(\varphi,A,\lambda\right)=S\left(\varphi,A\right) +
\int{\mathrm{d}}^2x
\sqrt{\gamma}\widehat\epsilon^{\mu\nu}\left(
-2\Gamma_a A^a_\nu\dd_\mu\lambda
-2\lambda A_\nu^a\dd_i\Gamma_a\dd_\mu\varphi^i
-\lambda f^a_{bc}\Gamma_aA^b_\mu A^c_\nu
\right)
\,\,.
\ee
The second and third terms of the Lagrange multiplier term can be
simply absorbed by shifting both $c_{ai}$ and $d_{ab}$
in the following manner
\be
c_{ai}\,\,\longrightarrow\,\, c_{ai}-\lambda\dd_i\Gamma_a
\,\,\,,\,\,\,
d_{ab}\,\,\longrightarrow\,\, d_{ab} -
\lambda f^c_{ab}\Gamma_c\,\,\,.
\label{shift}
\ee
This shift maintains the necessary conditions for gauge invariance
given in (\ref{gaugecond}).
Hence the only term in the Lagrange multiplier term which
is a genuine Lagrange multiplier is
\be
I\left(\varphi,A,\lambda\right)=\widetilde
S\left(\varphi,A\right) -2
\int{\mathrm{d}}^2x
\sqrt{\gamma}\widehat\epsilon^{\mu\nu}
\Gamma_a A^a_\nu\dd\lambda_\mu
\,\,\,,
\ee
where $\widetilde S(\varphi,A)$ is obtained from $S(\varphi,A)$ by
replacing $c_{ai}$ and $d_{ab}$ by their shifted values (\ref{shift}).
\par
Written in this form, the dual model has a global $U(1)$ symmetry
acting on $\lambda$ only
\be
\lambda\,\,\longrightarrow\,\, \lambda + \omega\,\,\,
\ee
with $\omega$ a constant. This transformation must be accompanied
by the corresponding shifts in $c_{ai}$ and $d_{ab}$.
It can be easily shown that the freedom in shifting these
two quantities stems from a symmetry in the defining equation
of $L_{ai}$ in (\ref{L}), namely $L_{ai}\,\longrightarrow\,
L_{ai} + \dd_i V_a$.
\par
If we now make an Abelian duality
with respect to this symmetry we recover the original model.
Gauging this $U(1)$ symmetry via the introduction of a $U(1)$
gauge field $B_\mu$
and adding a second Lagrange
multiplier $\kappa$  we obtain
\be
I\left(\varphi,\lambda, \kappa,A,B\right)
=
S\left(\varphi,A,\lambda\right) +
2\int{\mathrm{d}}^2x
\sqrt{\gamma}\widehat\epsilon^{\mu\nu}\left(
-\Gamma_a A^a_\nu B_\mu + \kappa\dd_\mu B_\nu\right)
\,\,\,.
\ee
Integrating out $B_\mu$ gives
\be
\Gamma_a A^a_\mu = \dd_\mu\kappa\,\,\,.
\ee
A gauge choice can now be made to obtain $A^a_\mu$ which would
lead to $F^a_{\mu\nu}=0$ and the original model is thus recovered.
Therefore in this method the dual of the dual theory
is the original theory as in the Abelian case.
\par
The procedure for eliminating the gauge fields from (\ref{newlag})
follows steps similar to those in section two. The dual action is obtained
by simply replacing in (\ref{dualaction}) $\lambda_a$ by
$\psi_a=\lambda\Gamma_a$.  The dual theory is
then obtained by eliminating the extra degrees
of freedom by gauge fixing. This procedure gives the dual
target space metric and the dual anti-symmetric tensor field.
The dual dilaton field is then given by the original dilaton
field plus the contributions arising from a proper consideration
of the path integral. Further details will be given elsewhere
\cite{me2}.

\section{Conclusions}
We have shown how a direct generalisation of
the methods of Abelian duality to non-Abelian duality leads
to the generation of a new local symmetry. This symmetry was not
taken into account in previous works on non-Abelian duality.
Due to this new symmetry, the isometry group is automatically
promoted to a larger non-semi-simple group.
\par
The fact that the dual backgrounds found in previous works did
not satisfy the conformal invariance conditions
(the vanishing of the beta functions) is due to not having taken into
account this extra symmetry. After eliminating the gauge
fields in order to obtain the dual theory, the new algebra
no longer closes. This is a major issue when dealing with gauge
theories. A treatment \`a la Batatin-Vilkovisky
using the master equation might be
of help \cite{vilko}. This is certainly worth investigating.
\par
The method we propose does not allow  the Lagrange multipliers
to carry any internal indices. These internal indices are at the
heart of the extra symmetry. Furthermore it was shown that in this
method the dual of the dual theory is the original theory as
expected.
\par
Explicit and concrete examples where this new non-Abelian duality
is constructed will be reported elsewhere
\cite{me2}. Another issue that needs
clarifications regards the global properties of the dual theory.
The way we obtained the dual theory suggests that there must be
connections between the global properties of the anti-symmetric
tensor field and those of the  Lagrange multipliers.
Finally it would be desirable
to see whether this duality could be  explained in terms canonical
transformations \cite{lozano}.
\newline
{\bf{Acknowledgements:}} I would like to thank
Peter Hodges, Tim Morris and Douglas Ross for
discussions.


\begin{thebibliography}{99}
\bibitem{amit}
A. Giveon, M. Porrati and E. Rabinovici, Phys. Rep. {\bf 244} (1994) 77;
\newline
E. Alvarez, L. Alvarez-Gaum\'e and Y. Lozano, {\it An Introduction to
T-Duality in String Theory}, hep-th/9410237.
\bibitem{narain}
K. S. Narain, Phys. Lett. {\bf B169} (1986) 41;
\newline
K. S. Narain, M. H. Sarmadi and E. Witten, Nucl. Phys. {\bf B279}
(1987) 369.
\bibitem{thelot}
A. Sen, Phys. Lett. {\bf B271} (1991) 295;
\newline
S. F. Hassan and A. Sen, Nucl. Phys. {\bf B375} (1992) 103;
\newline
J. Maharana and and J. H. Schwarz, {\it Non Compact Symmetries in
String Theory}, Caltech preprint, CALT-68-1790 (1992);
\newline
A. Giveon and M. Ro\u{c}ek, Nucl. Phys.
{\bf B380} (1992) 128;
\newline
E. Kiritsis, Nucl. Phys. {\bf B405} (1993) 109;
\newline
A. Giveon and E. Kiritsis, Nucl. Phys. {\bf B411} (1994) 487;
\newline
G. Veneziano, Phys. Lett. {\bf B265} (1991);
\newline
A.A. Tseytlin, Mod. Phys. Lett. {\bf A6} (1991) 1721;
\newline
K. A. Meissner and G. veneziano, Phys. Lett. {\bf B267} (1991) 33;
Mod. Phys. Lett. {\bf A6} (1991) 3397;
\newline
M. Gasperini and G. Veneziano, Phys. Lett. {\bf B277} (1992);
\newline
A. S. Schwarz and A. A. Tseytlin, Nucl. Phys. {\bf 399} (1993) 691.
\bibitem{lozano}
E. Alvarez, L. Alvarez-Gaum\'e, J.L.F.
Barb\'on and Y. Lozano, Nucl. Phys. {\bf B415} (1994) 71;
\newline
Y. Lozano, {\it Non-Abelian Duality and Canonical transformations},
preprint PUPT-1532, hep-th/9503045;
\newline
T. Curtright and C. Zachos, {\it Canonical Nonabelian Dual Transformations
in Supersymmetric Field Theories}, hep-th/95021126;
Phys. Rev. {\bf D49} (1994) 5408.
\bibitem{verlinde}
M. Ro\u{c}ek and E. Verlinde, Nucl. Phys. {\bf 373} (1992) 630.
\bibitem{busher}

T. H. Buscher, Phys. Lett. {\bf B201} (1988) 466;
\newline
B. E. Fridling and A. Jevicki, Phys. Lett. {\bf B134} (1984) 70;
\newline
E. S. Fradkin and A. A. Tseytlin, Ann. Phys. {\bf B162} (1985) 31.

\bibitem{delaossa}
X. de la Ossa and F. Quevedo, Nucl. Phys.
{\bf B403} (1993) 377.

\bibitem{israel}
A. Giveon and M. Ro\u{c}ek, Nucl. Phys. {\bf B421} (1994) 173;
\newline
M. Gasperini, R. Ricci and G. Veneziano,
Phys. Lett. {\bf B319} (1993) 438;
\newline
S. Elitzur, A. Giveon, E. Rabinovici, A. Schwimmer and G. Veneziano,
{\it Remarks on Non-abelian duality}, hep-th/9409011;
\newline
C. Klimc\'{\i}k and P. Severa, {\it Dual Non-abelian
Duality and the Drinfeld Double}, hep-th/9502122;
\newline
O. Alvarez, {\it Classical Geometry and Target Space
Duality}, hep-th/9511024.
\bibitem{me1}
N. Mohammedi,  {\it Classical Duality in Gauge Theories},
Southampton preprint, SHEP 95/23, hep-th/9507055.
\bibitem{ian}
I. Jack, D. R. T. Jones, N. Mohammedi and H. Osborn,
Nucl. Phys. {\bf 332} (1990) 359.
\bibitem{chris}
C. M. Hull and B. Spence, Phys. Lett. {\bf B232}
(1989) 204.
\bibitem{vilko}
I. A. Batalin and G. A. Vilkovisky, Phys. Lett. {\bf B102}
(1981) 27; Phys. Rev. {\bf D30} (1983) 2567.
\bibitem{me2}
N. Mohammedi, to appear.
\end{thebibliography}
\end{document}